\documentclass[
aps,
prl,
showpacs,
twocolumn,
superscriptaddress,
floatfix,
10pt
]
{revtex4-2}

\usepackage{amsmath,amssymb,amsfonts,stmaryrd,wasysym,graphicx,multirow,textcomp,subfigure}
\usepackage{url}
\usepackage[colorlinks=true,citecolor=blue,urlcolor=blue]{hyperref}
\usepackage{romannum}
\usepackage{mathtools}
\usepackage[utf8]{inputenc}
\usepackage{braket}
\usepackage{amsmath}
\usepackage{xcolor}
\usepackage{color,soul} 
\usepackage{bm}
\usepackage[normalem]{ulem}
\usepackage{amsmath}
\usepackage{epsfig}

\hypersetup{colorlinks=true, urlcolor=blue, citecolor=cyan, pdfborder={0 0 0}}
\newcommand{\yk}[1]{{\color{black}{{#1}}}}

\newcommand{\vv}[1]{{\boldsymbol #1}}

\pagecolor{white}

\begin{document}
\title{Tunable bulk photovoltaic effect in strained $\gamma$-GeSe}

\author{Hong-Guk Min}
\thanks{These authors contributed equally.}
\affiliation{Department of Physics, Sungkyunkwan University, Suwon 16419, Korea}
\author{Churlhi Lyi}
\thanks{These authors contributed equally.}
\affiliation{Department of Physics, Sungkyunkwan University, Suwon 16419, Korea}
\author{Youngkuk Kim}
\email{youngkuk@skku.edu}
\affiliation{Department of Physics, Sungkyunkwan University, Suwon 16419, Korea}
\date{\today}

\begin{abstract}
Recently, Lee \textit{et. al.} [Nano Lett. \textbf{21}, 4305 (2021)] synthesized monochalcogenide GeSe in a polar phase, referred to as $\gamma$-phase. Motivated by this work, we study the shift current of $\gamma$-GeSe and its tunability via an in-plane uniaxial strain. Using first-principles calculations, we uncover the electronic structure of the strained $\gamma$-GeSe systems. We then calculate frequency-dependent shift current conductivities at various strains. The tunability increases the shift current to $\sim$ 40 $\mu$A/V$^2$ \yk{for visible light}. Moreover, the direction of the shift current can be inverted by a light strain. Markedly, \yk{noticeable} behavior is found in the zero-frequency limit, which can be indicative of band inversion and electronic phase transition driven by the strain. Our results suggest that shift current can be tangible proof of bulk electronic states of $\gamma$-GeSe.
\end{abstract}
\maketitle

\section{Introduction}
A dramatic consequence of quantum mechanics featured in light-matter interaction is the generation of photovoltaic current without frequency dependence under light illumination.   An electric field  $\vv E (\omega)$ with angular frequency $\omega$ can drive a second-order nonlinear response of electrons that leads to a direct current  
\begin{equation}
J_a = \sum_{b c}\sigma_{abc}E^b(\omega) E^c(-\omega),
\end{equation}
referred to as shift current \cite{fridkin1973, Belinicher1980, von1981theory, Baltz81p5590, cote2002rectification, PhysRevLett.109.116601}. 
Here, $\sigma_{abc}$ is a shift current conductivity tensor and $a$, $b$, and $c$ represent coordinate indices.  Since the early discovery in 70's  \cite{auston1972optical, glass1975excited, koch1976anomalous}, 
the bulk photovoltaic effect has attracted much attention due to its potential applications in solar cells. The bulk photovoltaic effects can generally arise from diverse intra- and inter-band processes, including shift current, injection current, nonlinear Hall effect, and second-order jerk current \cite{PhysRevB.100.064301}. Notably,  shift current has attracted renewed attention due to its connection to band topology \cite{nastos2010optical, young2012first, young2013prediction},  allowing for design principles for high efficiency of light conversion beyond the Shockley-Queisser limit of conventional p-n junctions  \cite{Shockley1961}. This has enabled exciting discoveries of bulk photovoltaic materials, including many oxides \cite{, zhang1992resonant, chuang1992optical, glass1995high, fridkin2001bulk, somma2014high, tan2016shift}, quantum wells \cite{bieler2007shift, bieler2005ultrafast}, semiconductors \cite{laman2005ultrafast, nagaosa_tokura2019}, organic crystals \cite{ogden1984bulk, vijayaraghavan2014bulk},  nanostructures \cite{kral2000photogalvanic,ichiki2005photovoltaic, pintilie2007short, qin2008high, pintilie2010complex,  nakamura2016spontaneous}, and two-dimensional systems \cite{cook2017design, zhang2019switchable, PhysRevLett.119.067402}.

Along with the recent developments of topological states of matter, \cite{kane2013topological, bansil2016colloquium}, the shift current has attracted renewed attention in view of band topology. It is turned out that the excited electrons can prove the topology and geometry of the conduction and valence bands \cite{Inti15p216806, morimoto2016topological, Tan16p237402, cook2017design, PhysRevX.10.041041}. This picture explains the generation of shift current as being due to the change of electronic polarizations during the band transition of electrons dressed by photons \cite{morimoto2016topological}. The relationship between polarization and shift current of electrons is further demonstrated in monochalcogenide monolayers \cite{PhysRevLett.119.067402, cook2017design}, in which optimized polarization magnifies the shift current up to $\sim$ 100 $\mu$A/V$^2$. Despite these outstanding explorations, topological aspects of shift current have remained largely unexplored to date, partly due to the lack of suitable topological platforms. Encouragingly, a recent experiment reports the synthesis of a new polar monochalcogenide system GeSe \cite{Lee2021, kim2021quasiparticle}, providing an opportunity to explore the shift current phenomena.

In this paper, we perform first-principles calculations to study the shift current response of strained $\gamma$-GeSe. We calculate the shift current conductivity tensors for various in-plane uniaxial tensile strains. We show that the shift current of $\gamma$-GeSe is highly tunable in magnitude and direction within the 5\,\% tensile strain. In particular, a significant enhancement is expected in a high-frequency near visible-light regime, captured in, for example, the $\sigma_{zxx}$-component of the shift current conductivity tensor. In addition, a \yk{noticeable change} of shift current is found from, \yk{for example,} $\sigma_{zyy}$ \yk{and $\sigma_{zxx}$} near zero frequency. Electronic energy bands and the density of states are calculated to explain these behaviors. We attribute the strain-sensitivity of shift current to band inversion captured in the changes of orbital characters.

\begin{figure}
    \centering
\includegraphics[width=0.50\textwidth]{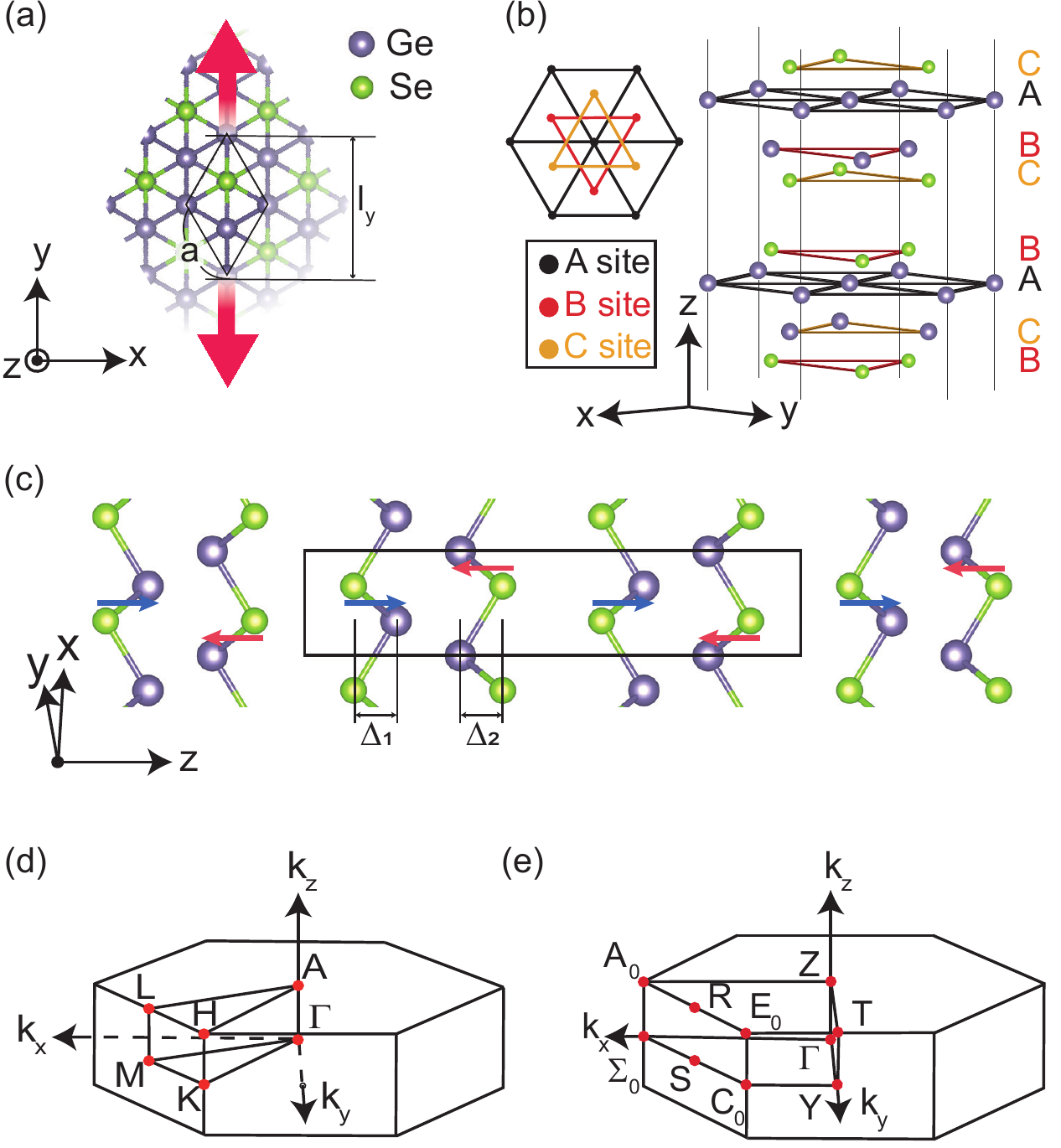}
\caption{
\label{fig:atomic_structure}
Crystal structure and the Brillouin zone (BZ) of $\gamma$-GeSe. (a) Top view of the pristine $\gamma$-GeSe atomic structure. Red arrows indicate the direction of a uniaxial strain. (b) Lateral view of the pristine $\gamma$-GeSe.  The $A$, $B$, and $C$ triangular sublattices in each Ge-Se honeycomb layer are indicated by black, red, and yellow color schemes, respectively. (c) Side view of the pristine $\gamma$-GeSe atomic structure. A primitive unit cell is indicated by a black box.  The buckling parameter $\Delta_1$ and $\Delta_2$ are introduced. The blue and red arrows indicate the polar direction of each layer. (d) Hexagonal BZ and high-symmetry points of the pristine $\gamma$-GeSe. (e) Orthorhombic BZ and corresponding high-symmetry points of the strained $\gamma$-GeSe.}
\end{figure}

\begin{figure*}
\includegraphics[width=0.90\textwidth]{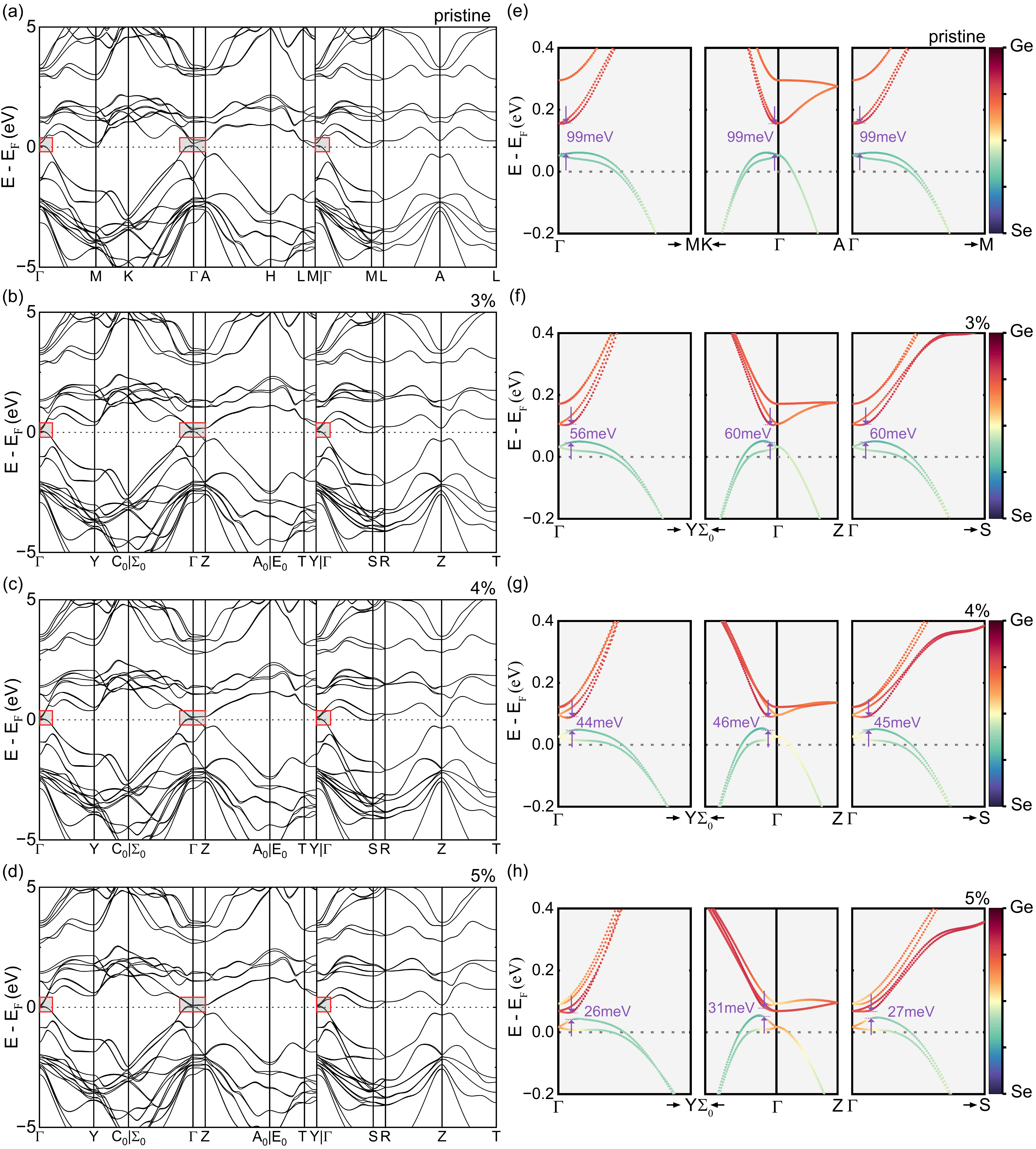}
\caption{
\label{fig:band_structure}
The band structure of $\gamma$-GeSe under a tensile uniaxial strain. (a) - (d) Energy bands of the pristine, 3\,\%, 4\,\%, and 5\,\% strained systems. (e) - (h) Magnified views of the boxed regions in the full energy bands of the pristine, 3\,\%, 4\,\%, and 5\,\% strained systems. The color coding represents the Ge and Se $s+p$ orbitals.}
\end{figure*}

\section{Computation details} 
Our first-principles calculations were performed based on density functional theory (DFT) as implemented in the \textsc{Quantum Espresso} package \cite{Giannozzi2009}. We employed the fully relativistic atomic pseudopotentials provided in the PseudoDojo library \cite{van_setten_pseudodojo_2018}. The employed pseudopotentials were built using the norm-conserving optimized Vanderbilt pseudopotential scheme \cite{hamann_optimized_2013}.
For describing the exchange-correlation energies, we used the Perdew-Burke-Ernzerhof generalized gradient approximation (PBE-GGA) \cite{perdew1996generalized}.
The plane wave basis was generated under the kinetic energy cutoff of 100 Ry.  Self-consistent fields were achieved by employing the Monkhorst-Pack sampling of $\vv k$ points from the $13 \times 13 \times 3$-grid of the first Brillouin zone (BZ)  \cite{monkhorst_special_1976}. The crystal structure of pristine $\gamma$-GeSe is fully relaxed with the $13 \times 13 \times 3$ Monkhorst-Pack $\vv k$-grid until the total force is fully converged below $10^{-4}$ Ry/Bohr. The fully relaxed lattice constants were calculated as 
$a \simeq 3.76$
\AA\,\,and $c \simeq 15.15$ \AA, which are in good agreement with the experiment \cite{Lee2021}. Under strain, we fixed the lattice constant only along the strained axis and relaxed the internal atomic coordinates and the other cell parameters. To calculate shift current, we used the Wannier-interpolation scheme \cite{PhysRevB.97.245143}. The Wannier Hamiltonians were first generated using the \textsc{Wannier90} package \cite{Mostofi2014, PhysRevB.97.245143}.  
To obtain an initial guess for the projection functions, we used the unified method for Wannier localization \cite{doi:10.1137/17M1129696, vitale_automated_2020}.
These Wannier Hamiltonians well reproduced the DFT bands in the $E < 10$ eV with respect to the Fermi level $E = 0$ eV. Using the constructed Wannier Hamiltonians, we calculated the shift current conductivity tensors using the \textsc{PostW90} package \cite{PhysRevB.97.245143}. The shift current conductivity tensors were calculated on the uniform $100 \times 100 \times 25$ $\vv k$-grid of the BZ. 
The $\mathbb{Z}_2$ topological invariants \cite{fu2007topological, moore2007topological} on time-reversal invariant planes and mirror Chern numbers \cite{teo2008surface, hsieh2012topological} on a $g_y$-invariant plane $k_y = 0$ were evaluated from the Wannier center flows \cite{soluyanov2011computing} as implemented in \textsc{WannierTools} \cite{WU2017}.

\section{Results and Discussion}
\subsection{Atomic structures and symmetries}
\label{section:atomic_str}
We begin by delineating the atomic structure and symmetries of $\gamma$-GeSe. As illustrated in Fig.\,\ref{fig:atomic_structure}, the unit cell of pristine $\gamma$-GeSe comprises eight atoms with four formula units. Each atom constitutes a distinct triangular sublattice layer, leading to eight vertically-stacked sublattices. The layered structure comprises two quadruple layers: $B$-$C$-$A$-$B$ and $C$-$B$-$A$-$C$ layers occupied Se-Ge-Ge-Se and Se-Ge-Ge-Se atoms, respectively. [See Fig.\,\ref{fig:atomic_structure}(b).].  The hexagonal BZ and corresponding high-symmetry $\vv k$-points of the pristine $\gamma$-GeSe are shown in Fig.\,\ref{fig:atomic_structure}(c). The pristine atomic structure preserves the hexagonal space group symmetries of $P 6_3 mc$ (\#186), which is generated by the mirror $M_{xy}$ and six-fold screw $S_{6z}=\{C_{6z}|00\tfrac{1}{2}\}$. Here, $C_{6z}$ is a six-fold rotation with respect to the $z$-axis, and $\{00\tfrac{1}{2}\}$ represents translation by half the out-of-plane ($z$-directional) unit vector. Spatial inversion is absent in $P6_3 mc$ (\#186), which is required to generate shift current. 

To reduce the crystalline symmetries, which can render qualitative changes in the electronic structure of $\gamma$-GeSe and correspondingly the shift current response, we consider a uniaxial tensile strain $\epsilon$, applied along the (110) armchair direction. As illustrated in Fig.\,\ref{fig:atomic_structure}(a), the uniaxial tensile strain $\epsilon$ is defined as $\epsilon = \frac{l_y - \sqrt 3 a} {\sqrt 3 a} \times 100 $\,\%, where $l_y$ is the stretched length of the unit cell along the $y$ direction. It is evident that the applied  strain lowers the six-fold screw symmetry $S_{6z} = \{C_{6z}|00\tfrac{1}{2}\}$ of the pristine structure into $S_{2z}=\{C_{2z}|00\tfrac{1}{2}\}$ of the strained one. Here, $C_{2z}$ is a two-fold rotation with respect to the $z$-axis. As a result, the strained $\gamma$-GeSe structure preserves the crystalline symmetries of the orthorhombic $Cmc 2_1$ (\#36) space group, generated by a glide mirror $g_y=\{M_y|00\tfrac{1}{2}\}$ and the two-fold screw $S_{2z}$.  The corresponding BZ and the high-symmetry $\vv k$-points are depicted in Fig.\,\ref{fig:atomic_structure}(d). 
We note that the presence of mirror $M_x$ and the glide $g_y$ is crucial to nullify the in-plane polarity in the strained system. By contrast, the out-of-plane polarization is allowed, attributed to the breaking of coplanar mirror symmetry by the buckling of Ge and Se sublattice in each Ge-Se layer. We find that applying tensile strains up to 5\,\% reduces the buckling parameter $\Delta_1$ and $\Delta_2$ by 3\,\%. Therefore, strains may significantly affect the buckling, which can be considered as a polar distortion, which is necessary to produce the shift current response.

\subsection{Electronic structures}
\label{section:elec_str}
\subsubsection{Band structures}
\label{subsection:bands_str}
Before presenting the shift current calculations, let us first clarify the electronic structure of pristine and strained $\gamma$-GeSe. Figures\,\ref{fig:band_structure}(a)-(d) shows the band structures of the pristine, 3, 4, and 5\,\%-strained systems, respectively.  A semimetallic nature is generically observed for the tested strains from the conduction and valence bands that overlap with the Fermi energy $E_{\rm F} = 0$ eV. Correspondingly, electron and hole pockets appear near the high symmetry $\Gamma$ and $M$ points, respectively. While the in-direct gap is absent, a close inspection reveals that a direct band gap exists throughout the entire BZ, which allows for the shift current calculations. For example, in the pristine case [Fig.\,\ref{fig:band_structure}(a)], the minimum direct band gap occurs in the vicinity of $\Gamma$ point, which is around 99 meV. Under a strain of up to 5\,\%, the semimetallic nature becomes more prominent. The size of the electron and the hole pockets increases with an increased overlap with the Fermi energy as shown in Figs.\,\ref{fig:band_structure}(a)-(d). The minimum direct band gap significantly decreases as the strain increases. From 0 to 5\,\%, the minimum direct band gap is monotonically reduced from 99 meV to 26 meV [See Figs.\,\ref{fig:band_structure}(e)-(f).].

The band inversion near the $\Gamma$ point is evident from the orbital projected bands shown in Figs.\,\ref{fig:band_structure}(e)-(h). For the pristine system, the valence and conduction bands mainly comprise the $sp^{3}$ orbitals of Se and Ge, respectively. The majority-orbital at $\Gamma$ starts to alter near the 4\,\% strain, signaling an inversion between the conduction and valence bands. The Se and Ge $sp^{3}$ orbitals acquire more portion in the 5\,\%-strained conduction and valance bands near $\Gamma$, respectively. The band inversion captured in the orbital-projected band structures is responsible for the abrupt changes of shift current near the zero frequency regime, as we will show later.

We observe from the calculated band structure that it captures the structural anisotropy of the layered system. The band structures on the $k_z = 0$ and $k_z = \pi$ planes resemble each other. This manifests the van der Waal-type weak interlayer interaction. Similarly, dispersion along $\Gamma$-$A$ is relatively weak. In addition, the valence bands exhibit a stronger dispersion than the rest of the bands, including the conduction bands. The enhanced dispersion of the valence bands can be attributed to the relatively stronger interlayer interaction within a quadruple layer than the inter-quadruple-layer interaction.

Noticeable changes are further found as follows. First, bands are less (more) dispersive along the parallel (normal) direction under strain. For example, the band dispersion is slightly increased (decreased) along \yk{$Y$}-$C_0$ ($\Gamma$-$Y$) direction. This captures the enhanced \yk{(dwindled) intra-layer} hopping of electrons under the \yk{in-plane uniaxial} tensile strain, which shrinks the \yk{in-plane perpendicular (colinear)} dimension and, correspondingly, the interlayer distances. In addition, the band degeneracy changes due to the reduced symmetries under strain. For example, the band crossings along the $\Gamma$-$A$ line open a gap due to the absence of $C_{3z}$. The space group $P6_3 mc$ (\#186) of the pristine system has three different two-dimensional irreducible representations $\Delta_7$, $\Delta_8$, and $\Delta_9$. In contrast, the $Cmc 2_1$ (\#36) space group of the strained system has only a single two-dimensional irreducible representation $\Delta_5$. Thus, the band crossings are absent, and only anti-crossings are present under strain along the high-symmetry $\Delta$ line.

\subsubsection{Density of states}
\label{subsection:dos}
\begin{figure*}[t!]
\centering \includegraphics[width=0.85\textwidth]{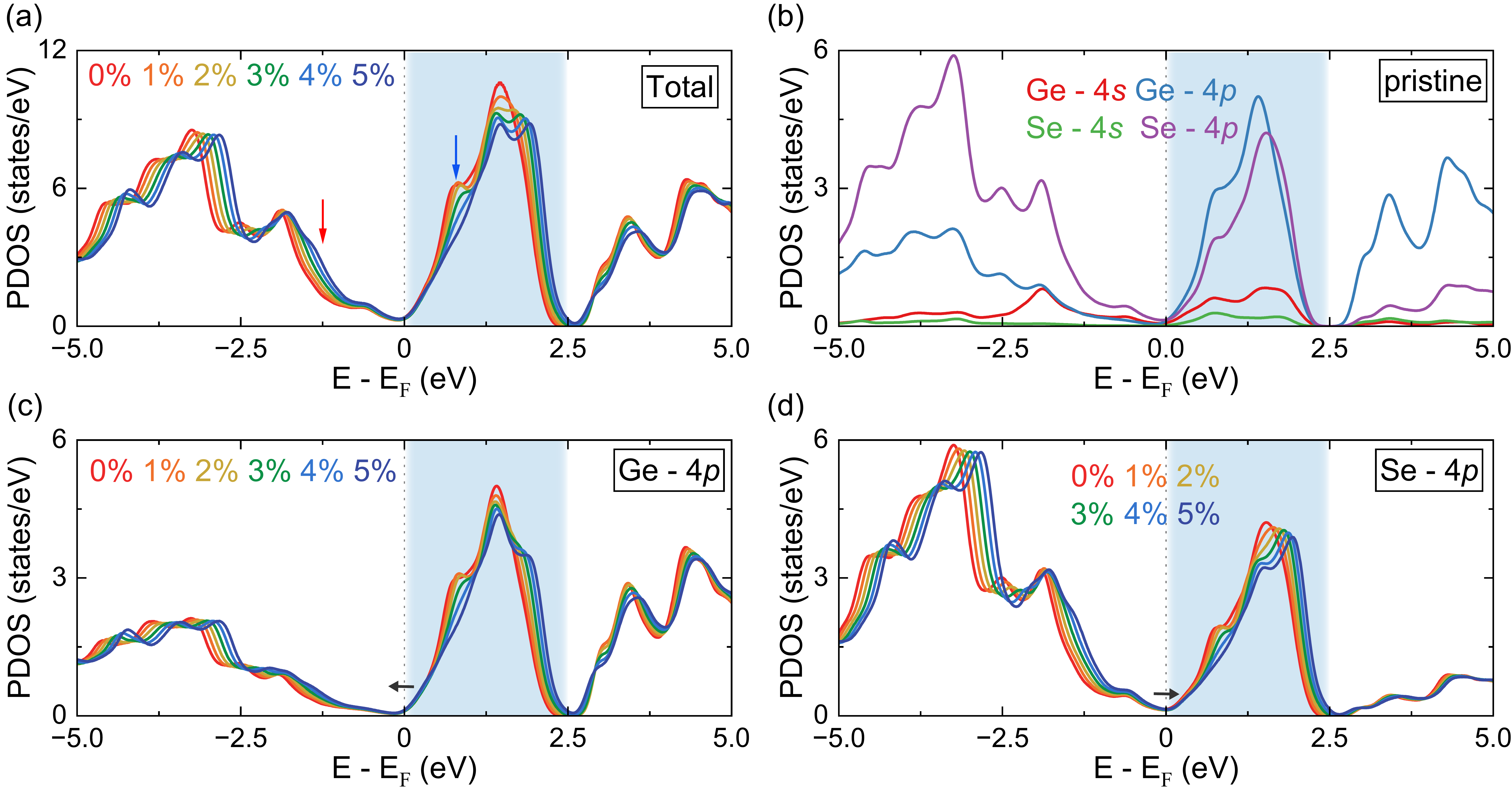}
\caption{\label{fig:dos}
The DOS and PDOS of $\gamma$-GeSe under a tensile strain. (a) A total density of states, obtained by the sum of the projected density of states. The change in the total DOS is shown by increasing the tensile strain from 0 to 5\,\%. 
The red and blue arrows indicate the peaks at -1.2 and 0.8 eV, respectively.
(b) Density of states, projected on the $4s$ and $4p$ orbitals of Ge and Se atoms for the pristine case. (c) $\sim$(f) The change of each projected density of states by increasing the tensile strain from 0 to 5\,\%. The black arrows indicate the direction of dip shifts as increasing the strain.
Blue boxes highlight the energy range in 0.0 eV $< E <$ 2.5 eV.  
}
\end{figure*}
The changes in electronic structure are more apparent in the total density of states (DOS) and the orbital-projected density of states (PDOS). Figures\,\ref{fig:dos} show the DFT DOS and PDOS. They signal the large shift current responses of $\gamma$-GeSe associated with the quasi-two-dimensional nature of the layered structure. In detail, intensified peaks appear throughout the -5.0 eV to 5.0 eV energy range. For example, considerable DOSs are accumulated in 0.0 eV $< E <$ 2.5 eV, in line with the formation of weakly dispersing bands in this energy range. This again captures the quasi-two-dimensional nature of $\gamma$-GeSe. The PDOSs in Figs.\,\ref{fig:dos}(a)-\ref{fig:dos}(d) show that these peaks consist of both Ge-4$p$ and Se-4$p$ orbitals. A close inspection reveals that the valence bands mainly comprise the Se-4$p$ orbitals, while the conduction bands consist of both Ge-4$p$ and Se-4$p$ orbitals. 

Examining detailed changes of total DOS driven by the strain, we observe the followings. First, the peaks are developed and suppressed at $E \simeq -1.2$ eV [red arrow in Fig.\,\ref{fig:dos}(a)] and $E \simeq 0.8$ eV [blue arrow in Fig.\,\ref{fig:dos}(a)], respectively, as we gradually increase the from 0 to 5\,\%. These trends are prominent in the PDOS of Ge and Se $4p$ orbitals. The suppression (development) of the peaks occurs at the conduction and valence bands due to the enhanced (weakened) hybridization of those orbitals by the strain. Second, there exists a blue shift 
of peaks and dips in the occupied energy regime, such as for $E \simeq$ -4.6, -3.8, -3.2, and -2.8. This matches well with the energy growth of the valance bands due to the tensile strain in Fig.\ref{fig:band_structure}. 
Similarly, a trend of blue shift appears when the strain suppresses the peaks of the conduction bands, as shown at $E \simeq$ 1.5,  3.4, and 4.3 eV. 
Finally, the Ge- and Se-4$p$ orbital characters are exchanged between the conduction and valence bands at the near zero dips at $E \sim 0$ as band inversion occurs [See the black arrows in Figs.\,\ref{fig:dos}(c) and (d)]. This trend is in line with the sensible dependence of shift current on the strain near zero energy, which we will discuss in detail in the next section. 

\subsection{Shift currents}
\label{section:shift_currents}

\begin{figure*}[t]
\includegraphics[width=0.9\textwidth]{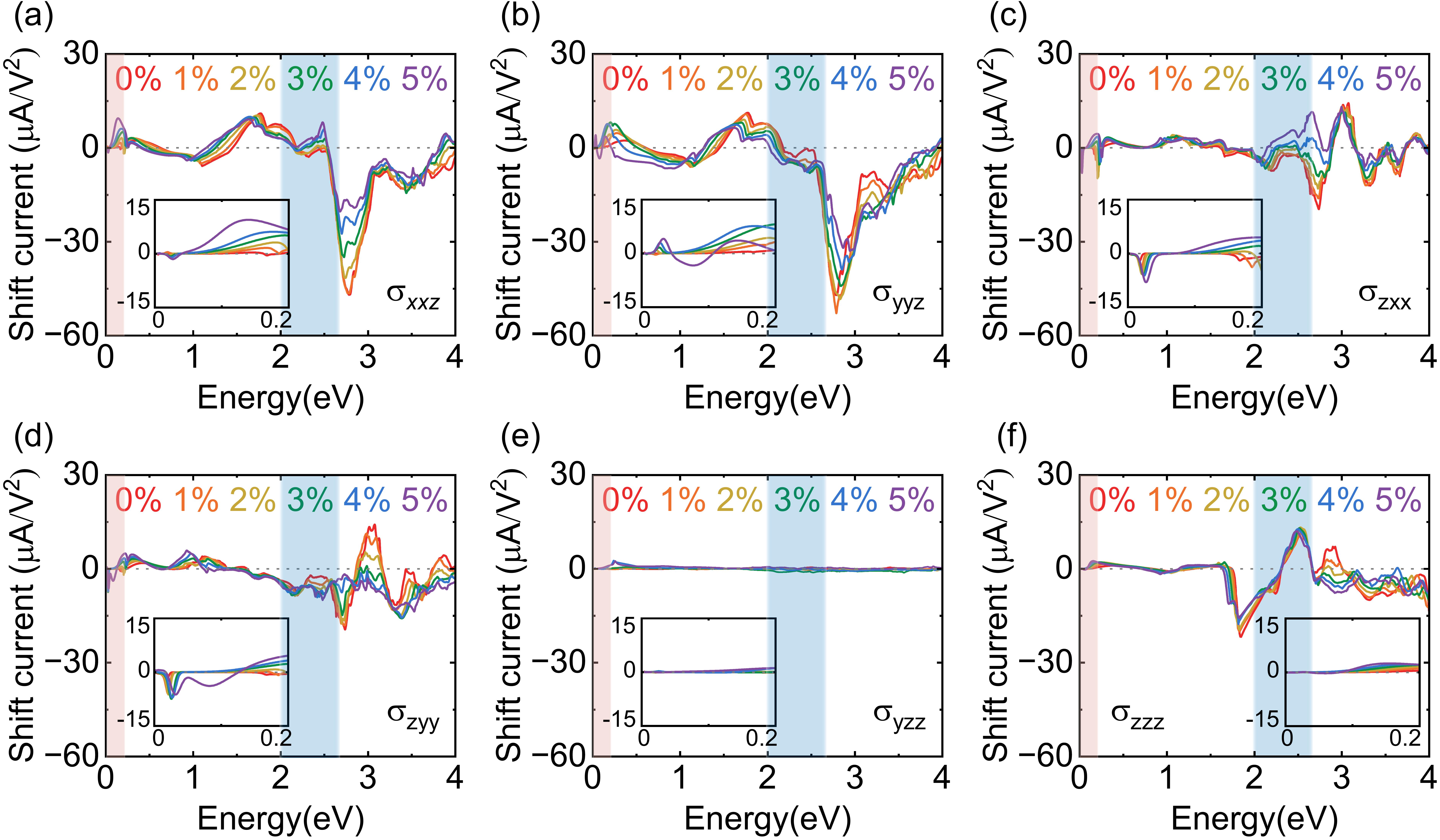}
\caption{\label{fig:sc} 
Shift current conductivity tensors of $\gamma$-GeSe as a function of uniaxial tensile strains. Different color schemes are used for different strains from 0 to 5\,\%, respectively. Blue boxes indicate the visible light regime from 2.0 to 2.7 eV. The red box highlight the energy regime near zero from 0 to 0.2 eV. \yk{The insets show a magnified view of the shift current for the 0 to 0.2 eV energies in the $\pm 15 \,\mu$A/V$^2$ shift current window.} 
}
\end{figure*}

Figure\,\ref{fig:sc} shows the shift current conductivity tensors calculated from our first-principles methods. Our calculations \yk{produce the shift current conductivity $\sigma_{abc}$ that is symmetric under the exchange of $b$ and $c$ indices in both pristine and strained systems}. This means that only eighteen components become distinct in the components of the three rank tensor. Out of these, \yk{only five components ($\sigma_{zxx}, \sigma_{zyy}, \sigma_{zzz}, \sigma_{yyz}$, $\sigma_{xxz}$) produce a non-negligible value, in good agreement with the symmetries of $P6_3mc$ ($Cmc2_1$) for the pristine (strained) $\Gamma$-GeSe \cite{gallego2019automatic}. Fig.\,\ref{fig:sc} shows the five non-vanishing components. In addition, as a representative example of a vanishing component, we show the calculation result of $\sigma_yzz$ in Fig.\,\ref{fig:sc}(e), which has a negligible value.} 

We further find a trend exists that those with an odd number of $z$ indices have a relatively larger value than those with an even number of $z$ indices. This also applies to the strained case, in which $C_{3z}$ is broken. The rationale for this as being due to \yk{symmetry constraints} arising associated with the $g_{x}$ and $M_{y}$ symmetries. An even number of $z$ components leads to an odd number of $x$ or $y$ indices. A non-zero value of these components infers that shift current can flow in a preferred direction indistinguishable by the system with $g_{x}$ and $M_{y}$. Thus, these components have a negligible value. Similarly, a symmetry constraint arises associated with the $C_{3z}$ symmetry manifested in the $xxz$ ($zxx$) and $yyz$ ($zyy$) components. They share a similar profile as a function of energy as shown in Figs.\,\ref{fig:sc}(a) and \ref{fig:sc}(b) [Figs.\,\ref{fig:sc}(c) and \ref{fig:sc}(d)]. Under strain, these components start to deviate, manifesting the absence of $C_{3z}$ symmetry.

Regarding the magnitude, we note that a considerable value of shift current is expected from $\gamma$-GeSe up to 
$\sim$ 40 $\mu$A/V$^2$ \yk{for visible light}. This value exceeds those of the known shift current materials.
For example, PbTiO$_3$ realized in a three-dimensional bulk has the maximum shift current conductivity tensor $\sim 10 \,\mu$A/V$^2$, BaTiO$_3$ with $18 \,\mu$A/V$^2$, and SbSI with $15 \,\mu$A/V$^2$ \cite{young_rappe_2012, nagaosa_tokura2019, fei_tan_rappe_2020}.
 
The strain affects the shift current response not only in its magnitude but also in its direction. In particular, the $zxx$-component exhibits the most prominent \yk{directional change} in response to the strain near the visible light frequency range (2.0 eV $< E <$ 2.7  eV) [See Fig.\,\ref{fig:sc}(c)]. One can observe a sign change in the current due to strain in this wide energy range. This sign change is important in particular, as it can be directly observed by the directional change of the current. The directional reversals are also observed in other components such as $\sigma_{xxz}$ and  $\sigma_{yyz}$ as shown in Figs.\,\ref{fig:sc}(a) and (b). We note that a directional reversal occurs within 4\,\%-strained, which is much more feasible than 5\,\% that induces band inversion. Such prominent changes in the shift current allow for an opportunity to engineer the direction of the photocurrent using the strain as a controllable knob. 

Finally, the shift current conductivity exhibits marked strain-dependence near zero-frequency fields. A few remarks and speculations are as follows.  First, \yk{the $yyz$-component is particularly enhanced with respect to $xxz$-component} upon applying 4\,\% strain, which is suppressed near the zero frequency. Therefore, the occurrence of the $y$-directional component upon {the exertion of an oblique incidence of a terahertz field} can be used as proof of crystalline symmetries. While a similar behavior also appears in the \yk{$zxx$ and $zyy$} components, a quantitative distinction exists between the \yk{$zxx$ and $zyy$} components, which can be proof of crystal axes. Last, the $zyy$ component shows an \yk{maxium rise near the 4\,\% strain for terahertz frequencies.} It \yk{appears} anomalous that shift current exhibits a non-zero value at the limit where $\omega = 0$. \yk{On close looking [See the inset of Fig.\,\ref{fig:sc}(d)], however, we find that the shift current response is smooth and well-behaved at zero energy. Nonetheless,}  we believe that the abrupt \yk{sink} of shift current can be experimental evidence of the semi-metallicity in the $\gamma$-GeSe system.

\begin{figure*}
\includegraphics[width=0.65\textwidth]{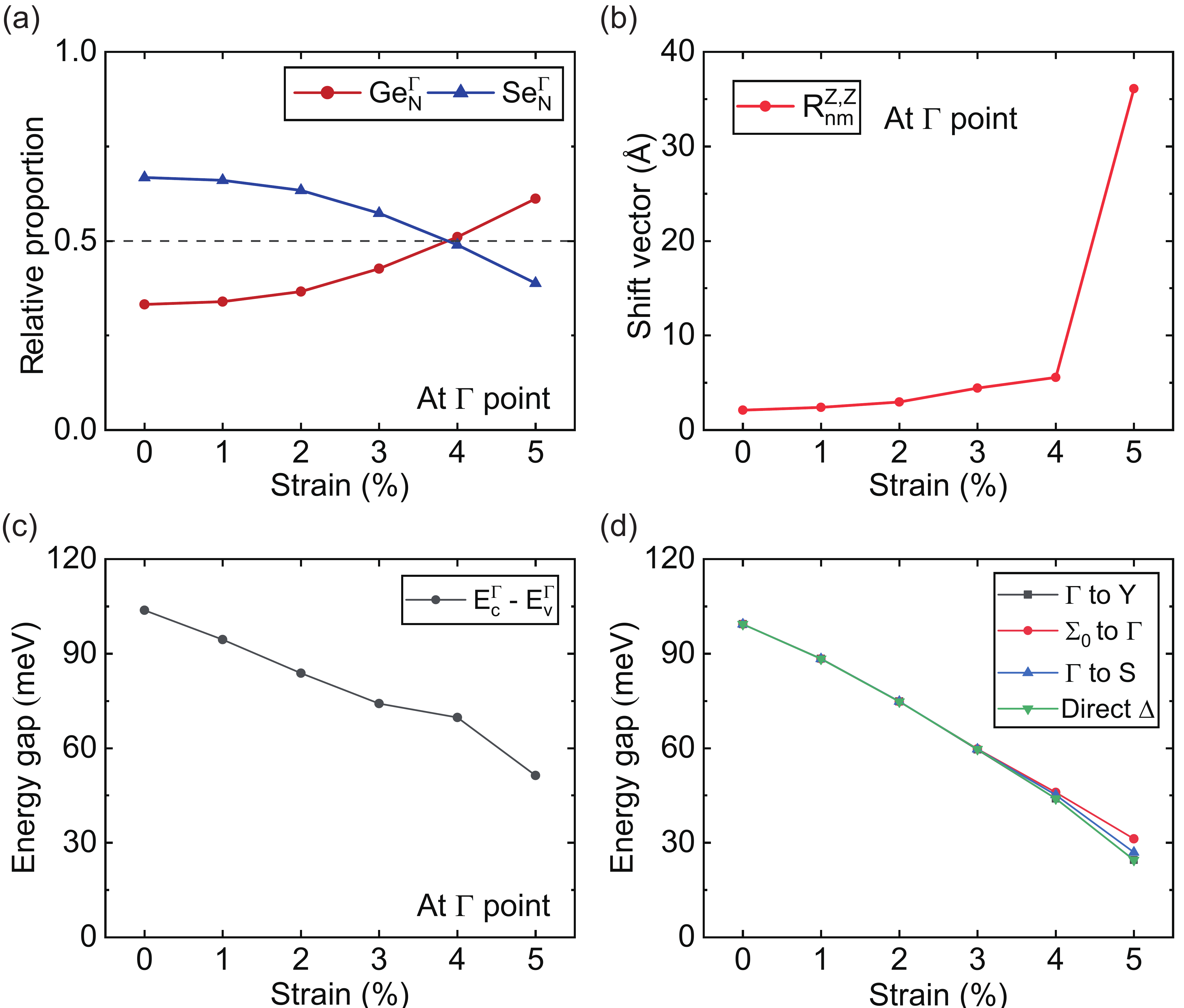}
\caption{
\label{fig:zero_sc}
(a) Relative proportion of Ge and Se $s$ and $p$ orbitals for the highest occupied ($N$-th) Bloch state at the $\Gamma$ point, where $N$ is electron filling. (b) $zz$-shift vector component contributed from $\Gamma$ point.
(c) Direct band gap at $\Gamma$ as a function of strain. (d) Minimum direct band gap on the high-symmetry lines.  
}
\end{figure*}

The abrupt change in shift current occurs near zero frequencies in close conjugation with the near zero-gap band inversion. The band inversion is featured in the orbital characters shown in Fig.\,\ref{fig:zero_sc}(a). The Se (Ge) $s+p$-orbital character becomes dominant beyond the 4\,\%\ strain in the highest occupied ($N$-th) Bloch state at $\Gamma$, where $N$ is electron filling. Fig.\,\ref{fig:zero_sc} shows that the band inversion significantly enhances the shift vector at $\Gamma$. We calculate the $zz$-component of the shift vector $R_{nm}^{z,z} = \frac{\partial \phi_{nm}^z}{\partial k_z} + A_{nn}^z - A_{mm}^z$ \cite{fregoso_morimoto_moore_2017}, where  $A_{nn}^z$ is the $z$-component of the Berry connection for the $n$-th band and $\frac{\partial \phi_{nm}^z}{\partial k_z}$ is the phase of the $z$-component of the interband transition dipole. This component of the shift vector is enhanced by 700\,\% from 5\,\AA\ to 35\,\AA\ via the band inversion. Thus, we speculate that the significant phase changes may accompany the orbital character change induced by band inversion. This, in turn, can \yk{be responsible for} an abrupt change in the shift current.

We attribute the origin of the considerable generation of shift current in $\gamma$-GeSe to the van der Waals layered geometry. As discussed in the band structure, DOS, and PDOS, the electronic structure of $\gamma$-GeSe consistently exhibits quasi-two-dimensional nature, which can also be featured in the joint density of states. Similar to the two-dimensional systems, in which photovoltaic current is enhanced due to the large joint density of states \cite{cook2017design, PhysRevLett.119.067402}, the layered geometry of $\gamma$-GeSe should play an essential role in generating significant shift current. In this respect, the van der Waal non-centrosymmetric systems are promising in generating large shift current. 

The shift current responses in these quasi-2D systems open exciting opportunities to study the dimensional crossover of shift current. The shift current along the out-of-plane direction is irrelevant in a genuine 2D system due to the vanishing shift vector responsible for the shift vector component.  By contrast, $\gamma$-GeSe in quasi-two dimensions generates a significant shift current along the out-of-plane $z$ direction. This observation motivates us to evaluate the shift vector components responsible for the $z$-directional shift current, such as the $zz$-component. Our explicit calculations of the shift vector reveal that the magnitude of the $zz$-component averaged over the whole BZ is $\sim$ 0.407\,\AA\ in the pristine layered $\gamma$-GeSe, while it readily reduces to $\sim$ 0.035\,\AA\ as we incorporate a 15\,\AA\ vacuum between the layers to suppress the interlayer coupling and mimic monolayer $\gamma$-GeSe. This calculation concludes that, whereas the joint DOS features a 2D nature, a dimensional crossover is yet to dominate the shift vector in pristine $\gamma$-GeSe, leading to the enhancement of shift current. Future studies should be required to pin down the role of dimensionality in the generation of shift current. Still, our results already demonstrate that strained $\gamma$-GeSe is an intriguing testbed for the dimensionality study.

The enlarged shift vector via band inversion resembles that of topological phase transitions \cite{Tan16p237402}. However, we point out that the exact topological phase transition is absent in this system within the 5\,\% strains. The absence of topological phase transition is first reflected in the band gap as a function of strain shown in Figs.\,\ref{fig:zero_sc}(c) and (d). Topological phase transition can occur via a band gap closer. However, the band gap remains open during the band inversion at $\Gamma$ and the high-symmetry lines. The absence of a gap closer indicates that these systems are adiabatically connected and, thus, topologically equivalent. Furthermore, our explicit calculations indeed confirm that time-reversal $\mathbb{Z}_2$ invariants \cite{fu2007topological, moore2007topological} and mirror Chern numbers \cite{teo2008surface, hsieh2012topological} are all trivial, in line with the band gap change. Thus, we conclude that although the band inversion induces a significant change in the polarization as dictated in the shift vector, topological states remain the same within the 5\,\% uniaxial strains.

Finally, we point out that our calculations have a clear limitation bound to single-particle approximation. Many-body effects can become prominent, in particular, in low-frequency regimes. We speculate that the Coulomb interaction may result in exciton binding between electrons and holes, enhancing the shift current response as in the monolayer GeS \cite{chan2021giant} and monolayer MoS$_2$ \cite{PhysRevB.89.235410}. In addition, the electron-phonon coupling may have a clear impact on shift current with sensitive dependence on the structural changes via strains, as demonstrated in BaTiO$_3$ \cite{PhysRevLett.126.177403}. We believe these effects should be included for high-fidelity prediction.

\section{Conclusion}
In conclusion, we have studied the shift current response of $\gamma$-GeSe under various uniaxial strains. We find a sensitive dependence of the shift current conductivity on the strain interplaying with the crystalline symmetries.  Our calculations have demonstrated that shift current is a tunable proof of various material properties. For example, meaningful information about the crystalline symmetries and, thus, the in-plane anisotropy can be proved by selectively measuring the shift current susceptibilities. More importantly, the shift current conductivity of $\gamma$-GeSe can inform the band gap. Experimentally, $\gamma$-GeSe is suggested as a low-gap semiconductor in Ref.\,\cite{Lee2021}. However, their optical conductivity clearly shows the Drude peak in the direct current limit, which inarguably necessitates further explorations to pin down the electronic state. In this respect, our results should be useful. Encouragingly, a thin film of WSe$_2$ has been strained up to 7 \% using an atomic force microscope on  SiO$_2$/Si substrates pre-patterned with hole arrays \cite{zhang2016elastic}. Thus, we believe it is highly feasible to measure the abrupt increase of out-of-plane shift current upon the exertion of an in-plane slowly-varying field, which can serve as an indicator of the semimetallic state of $\gamma$-GeSe.

\begin{acknowledgments}
This work was supported by the Korean National Research Foundation (NRF) Basic Research Laboratory (NRF-2020R1A4A3079707) and the NRF Grant numbers (NRF-2021R1A2C1013871). The computational resource was provided by the Korea Institute of Science and Technology Information (KISTI) (KSC-2020-CRE-0108).
\end{acknowledgments}


%

\end{document}